\documentclass[aps,prl,twocolumn,showpacs,superscriptaddress,groupedaddress]{revtex4-1}
\usepackage{graphicx}

\usepackage{amssymb}
\usepackage{amsmath}
\usepackage{latexsym}
\usepackage{ulem}
\usepackage{color}
\usepackage{dcolumn}
\usepackage{subfigure}
\usepackage[T1]{fontenc}
\usepackage[utf8]{inputenc}
\usepackage[english,french]{babel}
\usepackage[colorlinks=true,linkcolor=blue,citecolor=blue]{hyperref} %
\usepackage{bm}        % for math
\usepackage{amssymb}   % for math

\usepackage[capitalize]{cleveref}

\begin{document}

\title{Flat band induced room temperature ferromagnetism in two dimensional systems}

\author{G. Bouzerar}
\email[E-mail:]{georges.bouzerar@neel.cnrs.fr}
\affiliation{Université Grenoble Alpes, CNRS, Institut NEEL, F-38042 Grenoble, France}                    
\date{\today}
%\clearpage
\selectlanguage{english}
\begin{abstract}
The remarkable and fascinating properties of two-dimensional materials have raised them to the rank of most promising candidates for technological applications. In particular, the possibility of long-range ferromagnetic order in 2D materials is generating a growing excitement. Here, we demonstrate that flat bands (FBs) may pave the way to room temperature ferromagnetism in 2D compounds. Indeed, the magnetic exchanges between localized spins are largely dominated by the FB-FB contribution. This contribution is ferromagnetic and scales linearly with the local coupling, thus leading to cross-over temperatures (ferromagnetic phase/ paramagnetic phase) higher by an order of magnitude than those currently reported in experiments. Cross-over temperatures as high as $600~ K$ could be reached in micrometer-sized FB systems.
\end{abstract}
\pacs{75.50.Pp, 75.10.-b, 75.30.-m}

\maketitle
Since graphene has been discovered, the interest in two dimensional material (2DM) has grown tremendously. 
In the plethora of studies, two areas of research have particularly stood out: flat band (FB) physics, and magnetism in 2DM. In FB systems, the dispersionless bands are at the origin of unforeseen phenomena \cite{review1,review2} such as fractional quantum Hall states \cite{tang,neupert}, unconventional superconductivity \cite{miyahara,cao,yankowitz}, and magnetism \cite{lin,yin,tasaki}. FBs host as well an unusual type of quantum electronic transport as revealed in several studies \cite{gb-fb-paper1,mucciolo,vigh,qm-gb}. The interest for ferromagnetism in 2DM such as Cr$_2$Ge$_2$Te$_6$, CrI$_3$ or Fe$_3$GeTe$_2$ is experiencing a boost over the past years \cite{crgete,cri3,fegete,wang2D,yandong,miller,gibertini,santos}. 2DMs are undoubtedly promising candidates for technological applications, in spintronics, optoelectronics and data storage. Ferromagnetism in 2DMs was long ignored, because of Mermin-Wagner (MW) theorem \cite{mermin-wagner} that says that in one and two dimensions, continuous symmetries cannot be spontaneously broken at finite temperature in systems with short-range interactions. The MW theorem only excludes long-range magnetic order at finite temperature in the thermodynamic limit, i.e., for infinite systems. However, it does no exclude the possibility of quasi long-range ordering in finite systems below a cross-over temperature $T^{*}_{C}$ \cite{mag2D1,mag2D2,mag2D3}. It implies that below $T^{*}_{C}$ the spin-spin correlation length is much larger than the system size. So far, $T^{*}_{C}$ reported in micrometer-sized samples has been at most about $40\,K$ which stimulates the search for strategies to gain the missing order of magnitude.

\begin{figure}[t]\centerline
{\includegraphics[width=0.9\columnwidth,angle=0]{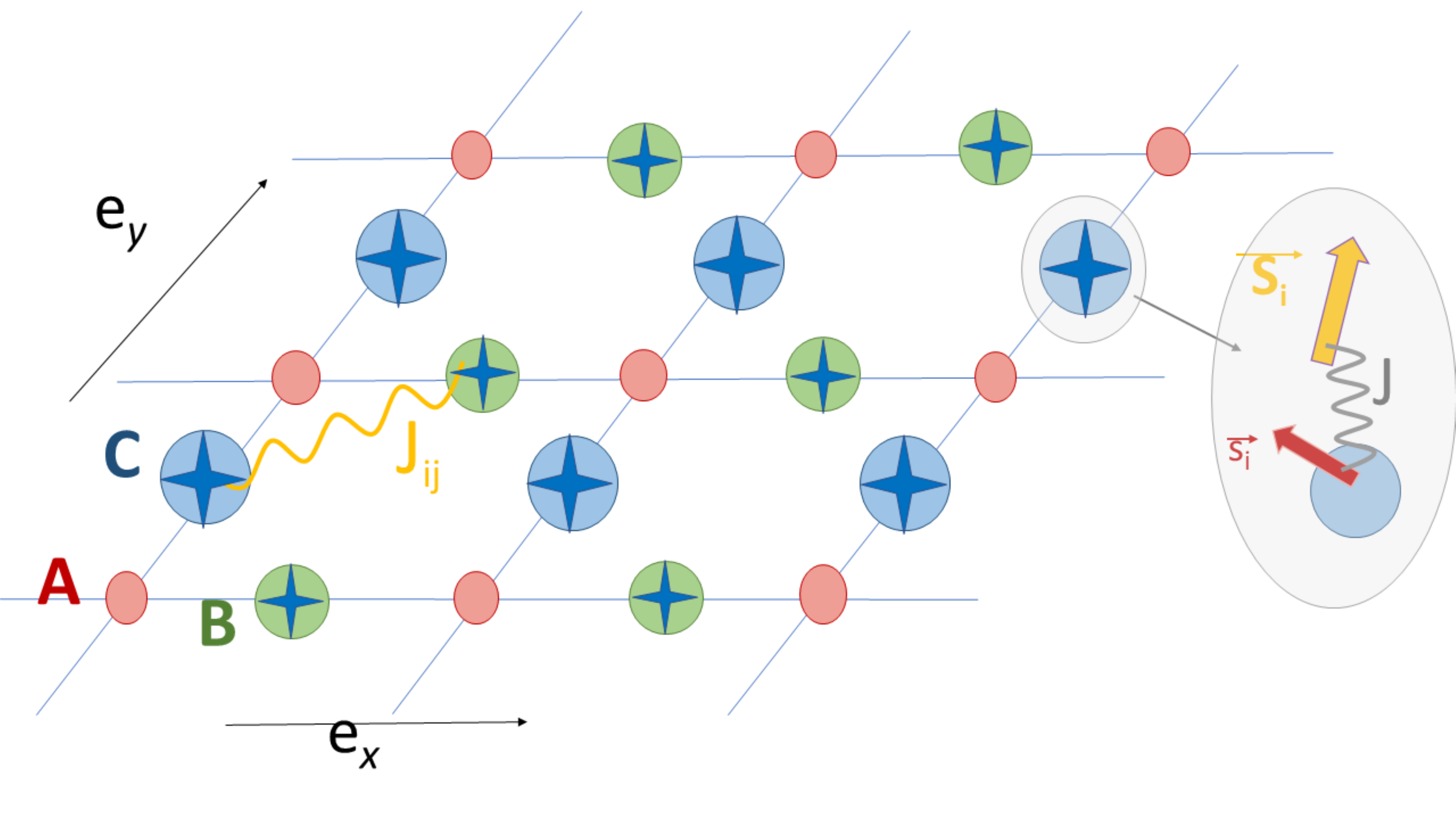}}
\vspace{-0.2cm} 
\caption{(Color online) Illustration of the Kondo Lieb Model. The hoppings are restricted to nearest neighbour pairs, crosses on B (C) sites correspond to the local exchange $J$ between the spin of the itinerant carrier ($\textbf{s}_i$) and the localized one $\textbf{S}_i$. $J_{ij}$ is the magnetic couplings between pairs of localized spins.}
\label{fig1}
\end{figure} 
Our aim is to discuss the impact of FBs on the magnetic properties of 2D systems. One key question is whether FBs may or may not promote ferromagnetism beyond room temperature. To address this issue, we consider the Kondo Lieb model as it is illustrated in Fig.~\ref{fig1}. The advantages of considering the Lieb lattice (LL) are numerous. First, the LL is a simple square lattice with 3 atoms/unit cell. It possesses a FB at $E=0$, there is no need to nano-structure or introduce defects or vacancies in the lattice. Finally, it could be designed experimentally in the framework of covalent-organic compounds for instance \cite{covorc}. The Hamiltonian reads,
\begin{eqnarray}
\widehat{H}=\sum_{\left\langle ij \right\rangle,\alpha} (t_{ij} c_{i\alpha}^{\dagger}c^{}_{j\alpha} + h.c.) + J\sum_{i \in B,C}\widehat{\bf s}_{i}\cdot {\bf S}_{i}.
\end{eqnarray}
c$_{i\alpha}^{\dagger}$ creates an electron with spin $\alpha=\uparrow,\downarrow$ at site $\textbf{R}_{i}$. In the first term, the sum runs over the lattice sites, $\left\langle ij \right\rangle$ are restricted to nearest neighbour pairs for which the hopping $t_{ij} = -t$. $J$ is the local Kondo coupling between the localized spin ${\bf S}_{i}$ at site $\textbf{R}_i$ and that of the itinerant carrier $\widehat{\bf s}_{i}$. Its components are $\widehat{s}^{\lambda}_{i} = c_{i\alpha}^{\dagger} \left[ {\widehat{\sigma}^{\lambda}}\right]_{\alpha\beta} c_{i\beta}$ where $\lambda=x,y,$ and $z$ and ${\widehat{\sigma}^{\lambda}}$ are the Pauli matrices. In what follows we set $t=1$ and $JS$ is expressed in units of $t$. In this study, we focus our attention on the half-filled case, thus the chemical potential $\mu=0$. 

The coupling between a pair of localized spins at $\textbf{R}_i$ and $\textbf{R}_j$ is given by \cite{jijcouplings},
\begin{eqnarray}
J_{ij}=-\frac{(JS)^{2}}{2\pi} \int_{-\infty}^{+\infty} \operatorname{\Im} \left[G_{ij}^{\uparrow}(\omega) G_{ji}^{\downarrow}(\omega) \right] f(\omega) d\omega. 
\label{eqcjij}
\end{eqnarray}
The Green's function $\widehat{G}^{\sigma}(\omega)=(\omega+i\eta-\widehat{H}^{\sigma})^{-1}$, where $\sigma=\uparrow,\downarrow$, $\eta$ mimics an infinitesimal inelastic scattering rate and $f(\omega)= \dfrac{1}{e^{\beta(\omega-\mu)}+1}$ is the Fermi-Dirac distribution. $J_{ij} \ge 0$ (resp. $J_{ij} \le 0$) means antiferromagnetic (resp. ferromagnetic) coupling. 

\begin{figure}[t]\centerline
{\includegraphics[width=0.9\columnwidth,angle=0]{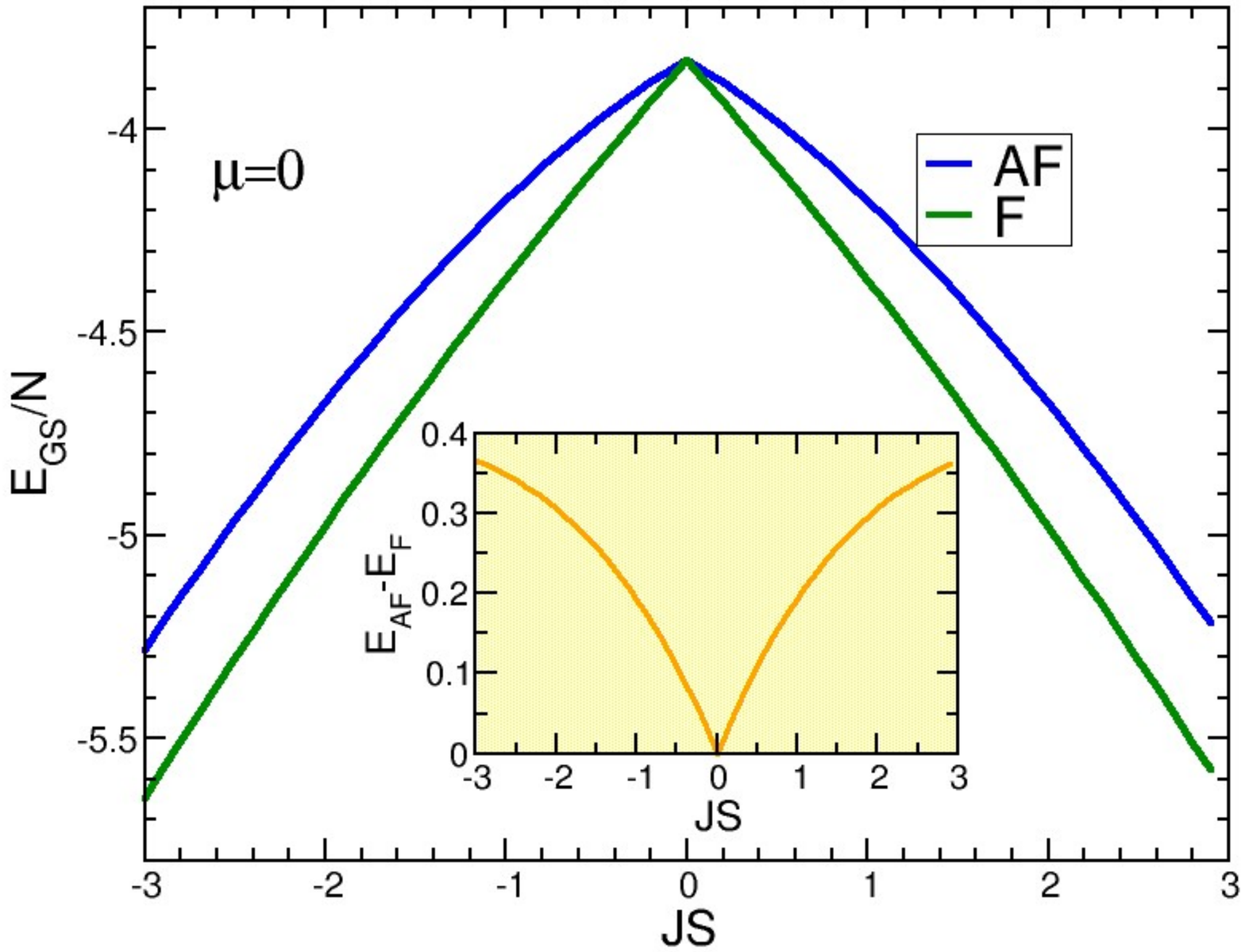}}
\vspace{-0.2cm} 
\caption{(Color online) 
Ground-state energy per unit cell $E_{GS}/N$ as a function of $JS$. $AF$ denotes the antiferromagnetic spin texture: $\textbf{S}_i=+S \textbf{e}_{z}$ (resp. $\textbf{S}_i=-S \textbf{e}_{z}$) on B (resp. C) sublattice. $F$ is the ferromagnetic spin texture, $\textbf{S}_i=+S \textbf{e}_{z}$ on both sublattices. The inset represents the energy difference between these two configurations.
}
\label{fig2}
\end{figure}

The calculation of $J_{ij}$ requires the knowledge of the ground-state (GS), hence that of the underlying localized spin texture. The spin configuration could be the disordered state (spins randomly oriented) corresponding to the high temperature phase, but we would have to deal with this issue numerically only. Instead, we consider the GS at $T=0\,K$ and restrict ourself to two different spin configurations: (i) the ferromagnetic GS (F-GS) where spins on B and C sublattices are parallel and (ii) the antiferromagnetic GS (AF-GS) where spins are anti-parallel. Fig.~\ref{fig2} shows the GS energy per unit cell $E_{GS}/N$ ($N$ is the number of unit cells) as a function of $JS$ for both spin configurations. First, as expected $E_{GS}/N$ is an even function of JS. Secondly, for $JS\ne 0$, F-GS has the lowest energy and the energy difference between these two spin configurations increases as $\vert JS \vert$ increases. 

\begin{figure}[t]\centerline
{\includegraphics[width=0.9\columnwidth,angle=0]{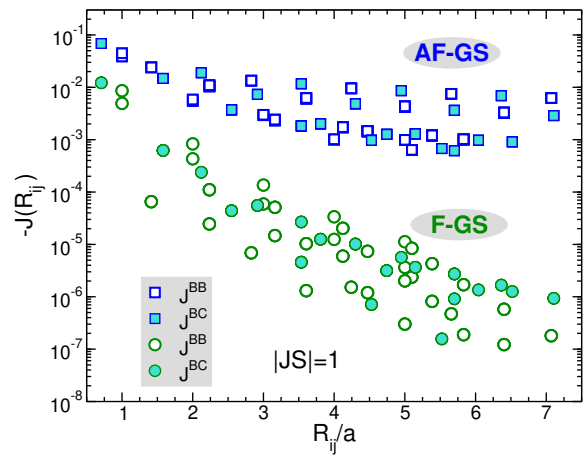}}
\vspace{-0.2cm} 
\caption{(Color online) 
Couplings (in units of $t$) as a function of the distance for $\vert JS \vert = t$.
The circles (resp. squares) correspond to the ferromagnetic (resp. antiferromagnetic) spin texture. The open (resp. filled) symbols correspond to $(B,B)$ (resp. $(B,C)$) couplings.
}
\label{fig3}
\end{figure}

Although F-GS has the lowest energy, we discuss the nature of the couplings for F-GS and AF-GS. The data are depicted in Fig.~\ref{fig3}. In both cases, the couplings are found always ferromagnetic. 
Finding ferromagnetic $(B,C)$ couplings for AF-GS is inconsistent with an antiferromagnetic ordering at $T=0 \,K$. Thus, even if we start with the wrong GS, the calculated couplings reveal the correct magnetic order.
Interestingly, the couplings are found significantly larger for AF-GS than for F-GS. More precisely, for short distances they are about 10 times larger in the first than in the second case, it becomes 3 orders of magnitude when $R_{ij} \ge 4\,a$.
The plotted data are obtained for $\vert JS \vert = t$ but our conclusions are general. From now on, we consider that F-GS is the GS at $T=0~K$.

\begin{figure*}
\centering
\includegraphics[width=0.9\textwidth]{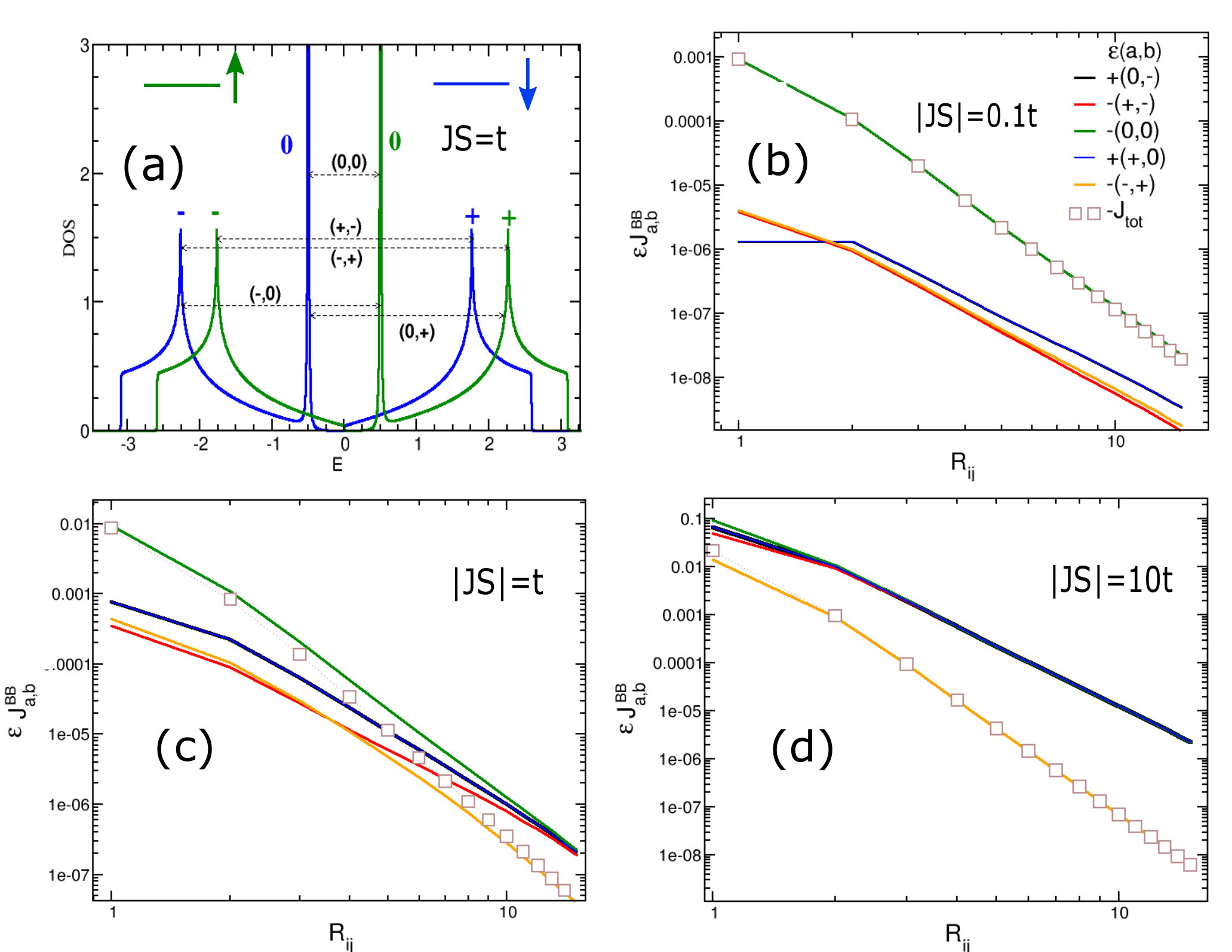}
\caption{
(a) Density of states as a function of the energy for $JS=t$.
(b-d) (B,B) coupling ($J_{tot}$) and its different contributions ($J_{a,b}$) along the $x-$ axis
as a function of the distance, for three different values of $\vert JS \vert/t$ ($0.1$, $1$ and $10$). The coefficient $\epsilon=1$ for antiferromagnetic couplings and $\epsilon=-1$ for ferromagnetic ones. $a$ (resp. $b$) is the band index in the $\uparrow$ (resp. $\downarrow$) spin-sector and $a$ and $b$ can be $-$ (lower dispersive band), $+$ (upper dispersive band) and $0$ (flat band).} 
\label{fig4}     
\end{figure*}

Now, we propose to discuss in details the nature of the couplings and their inter-band contributions. For that purpose, we first focus our attention on (B,B) pairs. For a given distance $R$, we write $J^{BB} = \sum_{a,b} J^{BB}_{a,b}$, where $a$ (resp. $b$) is the band index in the spin $\uparrow$-sector (resp. $\downarrow$-sector). $a$ and $b$ are $-$, $0$ and $+$, they correspond respectively to the lower dispersive band (DB), to the FB and to the upper DB as illustrated in Fig.~\ref{fig4}a. 
$J^{BB}$ along the $x$-axis and its different contributions are depicted in Fig.~\ref{fig4}b-d for 
$\vert JS \vert = 0.1$, $1$ and $10$. The five non vanishing terms correspond to the ($a$,$b$) pairs: $(-,0)$, $(-,+)$, $(0,0)$, $(+,-)$ and $(0,+)$. $J^{BB}_{-,0}$ and $J^{BB}_{0,+}$ are found antiferromagnetic and for symmetry reasons, identical. In contrast, the other three terms $J^{BB}_{-,+}$, $J^{BB}_{+,-}$ and $J^{BB}_{0,0}$ are ferromagnetic.
In the weak coupling regime ($\vert JS \vert= 0.1 \,t$), the dominant contribution to $J^{BB}$
is $J^{BB}_{0,0}$ which is orders of magnitude larger than the other contributions. A fit of the data plotted in Fig.~\ref{fig4}b, for $R/a \gg 1$, shows that $J^{BB} \propto 1/R^{\alpha}$ where $\alpha \approx 4$. In the intermediate regime ($\vert JS \vert = t$), the situation differs. For short distances $J^{BB}$ reduces to $J^{BB}_{0,0}$ whilst for larger ones it coincides with $J^{BB}_{+,-}$ and the sum $J^{BB}_{-,0} + J^{BB}_{-,+} + J^{BB}_{0,0} + J^{BB}_{0,+}$ vanishes. Finally, for $\vert JS \vert \gg t$, these four contributions cancel out each other for any distance and $J^{BB} \approx J^{BB}_{+,-}$ as illustrutated in Fig.~\ref{fig4}d. A fit of the data reveals that the couplings decay much more rapidly than in the weak coupling regime: $J^{BB} \propto 1/R^{\beta}$ where $\beta \approx 6$. These features are discussed in more details in the following. Notice as well, that our findings are general and valid for other directions and pairs of atoms.

To shed light on our numerical results, we go further and analytically derive the expression of the couplings in the weak and strong coupling regime. Details can be found in the Appendices A and B. Let us start with the weak coupling regime ($\vert JS \vert \ll t$) for which one needs to focus only on the FB-FB contribution. The FB eigenvalues and eigenstates for both spin sectors are, $E^{\sigma}_{0}= \pm \frac{JS}{2}=\pm \Delta$ where $+$ (resp.$-$) corresponds to $\sigma=\uparrow$ (resp. $\sigma=\downarrow$) and $\langle \Psi^{\sigma}_{0} \vert= (0,\frac{f_{y}}{s},-\frac{f_{x}}{s})$, where $f_{\lambda}=-2t\cos(k_{\lambda}a/2)$, with $\lambda=x,y$ and $s(\textbf{k})=\sqrt{f^{2}_x(\textbf{k}) + f^{2}_y(\textbf{k})}$. Starting from the definition of the couplings, we get,
\begin{eqnarray}
J^{XY}_{0,0}(\textbf{R})=-\frac{1}{2}\vert JS \vert \Big| \frac{1}{N}\sum_{\textbf{k}}\dfrac{f_{XY}(\textbf{k})}{s^{2}(\textbf{k})} e^{i{\textbf{k}.\textbf{R}}} \Big|^{2},
\label{J00b}
\end{eqnarray}
where $f_{XY}(\textbf{k})=f^{2}_y(\textbf{k})$, $f^{2}_x(\textbf{k})$ and $f_x(\textbf{k}).f_y(\textbf{k})$
for respectively $(X,Y)=(B,B)$, $(C,C)$ and $(B,C)$. 
Eq.(\ref{J00b}) shows that for any pairs of atoms, $J^{XY}_{0,0}(\textbf{R})$ is always ferromagnetic and varies linearly with $\vert JS \vert$. This contrasts with the standard weak coupling regime which leads to a $(JS)^{2}$ dependency of the couplings. It simply means that the perturbative calculation breaks down in the presence of FBs. Indeed, the FB-FB contribution is simply absent in the perturbative calculation since the Green's functions used are those corresponding to $JS=0$. In a recent work, within such an approach, it has been shown that the FB-DB terms introduce strong frustration effects \cite{gb-couplings}. This is still partly correct, for $\vert JS \vert \ll t$, the FB-DB contributions are antiferromagnetic and scale as $(JS)^{2}$. However, $J^{XY}_{0,0}(\textbf{R})$ largely dominates and washes out completely the frustration effects.
This explains as well why in the numerical data plotted in Fig.~\ref{fig4}b, the FB-FB term is much larger that the other contributions by at least one order of magnitude. The complete analytical calculations as detailed in Appendix A lead to,
\begin{eqnarray}
J^{BB}_{0,0}(\textbf{R})=J^{CC}_{0,0}(\textbf{R})=-\vert JS \vert \dfrac{a^{4}}{8\pi^{2}R^{4}} \cos^{2}(2\theta),
\label{jbbanalyt}
\\
J^{BC}_{0,0}(\textbf{R})=-\vert JS \vert \dfrac{a^{4}}{8\pi^{2}R^{4}} \sin^{2}(2\theta).
\end{eqnarray}
$\theta$ is the angle between $\textbf{R}$ and the $x$-axis. Eq.(\ref{jbbanalyt}) clearly explains the $1/R^{4}$ decay of the couplings found in the numerical calculations. Let us define the dimensionless quantity $C^{0}=\dfrac{J^{BB}_{0,0}}{\vert JS \vert}\cdot\left[\frac{R}{a}\right]^{4}$. A fit of the numerical data plotted in Fig.\ref{fig4}(b) gives $C^{0} = -0.0123$ which agrees very well with the analytical result $C^{0} = - 0.0126$.

We now switch to the strong coupling regime ($\vert JS \vert \gg t$) for which the couplings reduce
to $J^{XY}_{-,+}$. The energy of the DBs in each spin sector are $E^{\uparrow}_{\pm}= -g_{\mp}$ and $E^{\downarrow}_{\pm}=g_{\pm}$ where $g_{\pm}=\frac{1}{2} \left[-\Delta \pm \sqrt{\Delta^{2}+4s^{2}} \right]$. The corresponding eigenvectors are respectively, $\langle \Psi^{\uparrow}_{\pm} \vert= \frac{1}{D_{\pm}} (g_{\pm},f_{x},f_{y})$ and $\langle \Psi^{\downarrow}_{\pm} \vert= \frac{1}{D_{\mp}}
(-g_{\mp},f_{x},f_{y})$, where $D_{\pm}=\sqrt{s^{2}+ g_{\pm}^{2}}$. Because, the full calculation of
$J^{XY}_{-,+}$ is lengthy, the details can be found in the appendix B. For $\vert JS \vert \gg t$, we find,
\begin{eqnarray}
J^{BB}_{-,+}(\textbf{R})=-\frac{1}{\pi^{2}} (4\cos^{2}(\theta)-1)^{2} \frac{a^{6}}{R^{6}}\frac{t^{2}}{\vert JS \vert},
\label{jbb-jsinfty-b}
\\
J^{BC}_{-,+}(\textbf{R})=-\frac{3}{\pi^{2}} \sin^{2}(2\theta) \frac{a^{6}}{R^{6}} \frac{t^{2}}{\vert JS \vert}.
\label{jbc-jsinfty-b}
\end{eqnarray}
Notice that $J^{CC}_{-,+}(\textbf{R})$ is straightforwardly obtained by replacing in $J^{BB}_{-,+}$ $\theta$ by $\frac{\pi}{2} + \theta $. These expressions clearly explain the $1/R^{6}$ decay of the (B,B) couplings observed in the numerical calculations. As before, we define the dimensionless variable $C^{\infty}=J^{BB}_{-,+}\cdot\left[\frac{R}{a}\right]^{6}\cdot\frac{\vert JS \vert}{t^{2}}$. From a fit of the numerical data plotted in Fig.\ref{fig4}(c) we obtain $C^{\infty} = -0.71$, whilst Eq.(\ref{jbb-jsinfty-b}) gives for $\theta=0$, $C^{\infty}=-0.91$. The agreement between analytical and numerical calculations is not as good as that found for $\vert JS \vert \ll t$. However, for other directions, the agreement is much better. For $\theta= \frac{\pi}{2}$, Eq.(\ref{jbb-jsinfty-b}) gives $C^{\infty}=-0.101$ and from the numerical data one gets $C^{\infty}=-0.102$. On the other hand, for $J^{BC}_{-,+}$ an excellent agreement is found for any values of $\theta$. More details are available in Appendix B. 

In order to derive the cross-over temperature $T^{\star}_C$, we now turn to the calculation of the magnetic excitations. It is important to point out that, because the couplings decay sufficiently rapidly, in other words $\sum_{\textbf{R}}J^{XY}(\textbf{R})R^{2}$ is finite for any $(X,Y)$ pair, the Mermin-Wagner theorem \cite{mermin-wagner} implies that in the thermodynamic limit (system size $\rightarrow \infty$) the Curie temperature ($T_C$) vanishes. A finite $T_C$ in 2DM is possible only in the presence of anisotropy, which opens a gap in the Goldstone mode. However, even in absence of anisotropy, for a finite system one can define a size dependent cross-over temperature $T^{\star}_C$. A relevant question is, what is the order of magnitude of $T^{\star}_C$ for a micrometer-sized sample which is typically that of samples realized in laboratories. An appropriate tool to estimate $T^{\star}_C$ is the Random Phase Approximation (RPA) \cite{tyablicov}. In diluted magnetic semiconductors and manganites, the comparison with Monte Carlo simulation has revealed that RPA is reliable and accurate \cite{LRPA1,LRPA2}. The Heisenberg Hamiltonian reads,
\begin{eqnarray}
H^{H}=\frac{1}{2}\sum_{i\ne j} \mathcal{J}_{ij} {\bf S}_{i}\cdot {\bf S}_{j},
\end{eqnarray}
where $\mathcal{J}_{ij}S^{2}=J_{ij}$. To calculate the magnetic properties, we use the equation of motion method applied to the retarded spin Green's function, $G^{S}_{ij,XY}(\omega)=\int^{+\infty}_{-\infty} G^{S}_{ij,XY}(t)e^{i\omega t}dt$ where $G^{S}_{ij,XY}(t)=-i\theta(t) \bigl \langle \left[ S_{X,i}^{+},S_{Y,j}^{-}\right] \bigr\rangle$ and, $\bigl \langle ... \bigr\rangle$ denotes the thermal average. We only present the main results, the full procedure is detailed in Appendix C. As expected, we find two magnon branches, $\omega_{\textbf{q}}^{\pm}=\bigl\langle S^{z} \bigr\rangle E_{\textbf{q}}^{\pm}$, where $-$' is the acoustic mode and $'+'$ is the optical one, and,
\begin{eqnarray}
E_{\textbf{q}}^{\pm}=f_{+}^{\textbf{q}} \pm \sqrt{(f^{\textbf{q}}_{-})^{2}+(f^{\textbf{q}}_{BC})^{2}},
\end{eqnarray}
where $f^{\textbf{q}}_{+}=\frac{1}{2} (f^{\textbf{q}}_{BB} + f^{\textbf{q}}_{CC})$ and $f^{\textbf{q}}_{-}=\frac{1}{2} (f^{\textbf{q}}_{BB} - f^{\textbf{q}}_{CC})$ with $f^{\textbf{q}}_{XX}=-\sum_{Y} \bar{\mathcal{J}}^{XY}(\textbf{0})
+\bar{\mathcal{J}}^{XX}(\textbf{q}) $ ($X=B,C$), we have defined as well $\bar{\mathcal{J}}^{XY}(\textbf{q})=\sum_{\textbf{R}} e^{i\textbf{q}.\textbf{R}} \mathcal{J}^{XY}(\textbf{R})$ and $f^{\textbf{q}}_{BC}=\bar{\mathcal{J}}^{BC}(\textbf{q})$.
As it is shown in Appendix C, the cross-over temperature is given by,
\begin{eqnarray}
k_{B}T^{\star}_{C}= \frac{1}{3}(1+\frac{1}{S})\left[ 
\frac{1}{N}\sum_{\textbf{q}\neq \textbf{0},\lambda=\pm} \dfrac{A^{\lambda}_{\textbf{q}}}{S^{2}E_{\textbf{q}}^{\lambda}} \right]^{-1}.
\label{eq-Tc}
\end{eqnarray}
$A^{\pm}_{\textbf{q}}$ is the spectral weight on each magnon branch. We can rewrite Eq.(\ref{eq-Tc}) $T^{\star}_{C}=(1+\frac{1}{S})T^{\star,cl}_{C}$, where $T^{\star,cl}_{C}$ is the cross-over temperature for classical spins ($S \rightarrow \infty$). 

Let us briefly discuss the size dependence of $T^{\star,cl}_{C}$. We consider square shaped flakes of size $La \times La$. 
Using the fact that for small $|\textbf{q}|$, $A^{\pm}_{\textbf{q}} \approx 1/2$, $E_{\textbf{q}}^{+} \approx E_{\textbf{0}}^{+}$ and $E_{\textbf{q}}^{-} \approx Dq^{2}$, where $D$ is the spin stiffness of the acoustic mode, we obtain, $\frac{1}{N}\sum_{\textbf{q}\neq \textbf{0},\lambda=\pm} \dfrac{A^{\lambda}_{\textbf{q}}}{S^{2}E_{\textbf{q}}^{\lambda}} \approx \frac{1}{2S^{2}} \left[ \frac{1}{E_{\textbf{0}}^{+}} + \frac{a^{2}}{2\pi D} \ln(\frac{Q_{C}}{q_{min}}) \right]$ where $Q_{C}$ is a cut-off, and $Q_{min}=\frac{2\pi}{La}$ is the smallest non zero momentum. Then Eq.(\ref{eq-Tc}) becomes, 
\begin{eqnarray}
k_{B}T^{\star,cl}_{C}= \dfrac{2S^{2}}{A+B\ln(\frac{L}{2\pi})}.
\label{Tc_log}
\end{eqnarray}
$A=\frac{1}{E_{\textbf{0}}^{+}} + \frac{a^{2}}{2\pi D} \ln({a.Q_{C}}) $ and $B= \frac{a^{2}}{2\pi D}$. 
As expected, $T^{\star,cl}_{C}$ decays slowly as $1/\ln(L)$, and in accordance with the MW theorem, it vanishes in the limit $L\rightarrow \infty$. In the presence of a small anisotropy leading to a gap ($\Delta$) opening, 
$\ln(\frac{L}{2\pi})$ is replaced by $\frac{1}{2}\ln(\frac{D}{\Delta a^{2}})$ and $T^{\star,cl}_{C}$ becomes the true Curie temperature.

\begin{figure}[t]\centerline
{\includegraphics[width=1.0\columnwidth,angle=0]{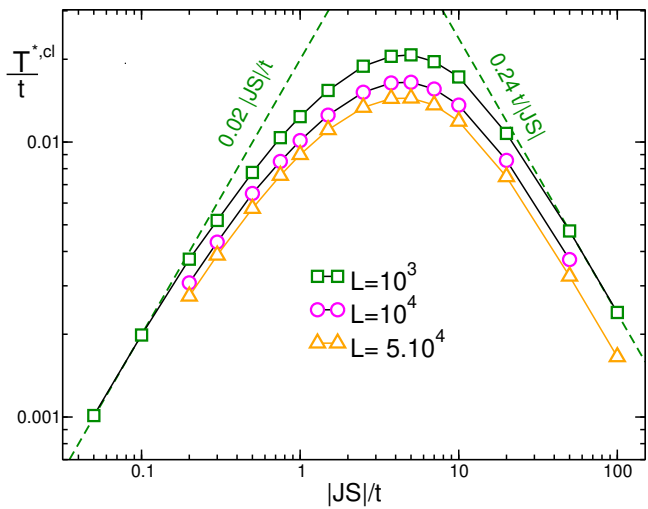}}
\vspace{-0.3cm}
\caption{(Color online) 
Cross-over temperature $T^{\star,cl}_{C}$ as a function of $\vert JS \vert$ for three different sizes $L=10^{3}$, $10^{4}$ and $5.~10^{4}$. The green dashed lines are fits for the weak and strong coupling regimes for the data obtained for $L=10^{3}$. 
} 
\label{fig5}
\end{figure} 

To illustrate the reliability of RPA, we compare our expression of $T^{*,cl}_C$ for the nearest neighbour Heisenberg Hamiltonian one the square lattice with that obtained from Monte Carlo simulations (MC) \cite{santos}.  
The system considered is a $1 \mu m \times 1\mu m$ flake with a lattice spacing $a = 4\mathring{A}$. Using Eq. (\ref{Tc_log}), we get $T^{\star,RPA}_{C} \approx 0.58~J$ whilst it has been found $T^{\star,MC}_{C} \approx 0.535~J$, where $J$ denotes the nearest neighbour coupling between the classical spins. We observe that the agreement between RPA and MC is surprisingly good.

We confidently return to our system. In Fig.\ref{fig5}, $T^{\star,cl}_{C}$, calculated numerically is plotted as a function of $\vert JS \vert$ for three different sizes, the biggest system contains $12.5$ billions of atoms.
As can be seen, an increase of $L$ has a small impact on $T^{\star,cl}_{C}$. This is consistent with the $\ln(L)$ dependency explicitly expressed in Eq.(\ref{Tc_log}). We now, estimate the cross-over temperature in mesoscopic samples. We consider a lattice spacing $a$ of 3$\mathring{A}$ and choose for $t$ a value of $1\,eV$, we recall that in graphene $t\approx 2.7 \,eV$. From Fig.\ref{fig5} the maximum of the cross-over temperature $T^{\star}_{max}$ in a $3 \mu m \times 3 \mu m$ sample is $612~K$, $410~K$ and $342~K$ for respectively $S=1/2$, $S=1$ and $S=3/2$.
These temperatures are beyond room temperature and one order of magnitude higher than those reported in various 2DMs such as Cr$_2$Ge$_2$Te$_6$ \cite{crgete} and Fe$_3$GeTe$_2$ \cite{fegete}.

In conclusion, FB states could pave the way to ambient ferromagnetism in 2D materials.
It is revealed, in the weak and intermediate coupling regime, that the magnetic exchanges between localized spins are largely dominated by the ferromagnetic FB-FB contribution which scales linearly with the local coupling, contrasting with the standard quadratic dependence of the other contributions. Using reasonable physical parameters, we find that cross-over temperatures well beyond $300~K$ could be reached in micrometer-sized systems opening interesting avenues towards technological applications.

%\begin{acknowledgments}
%\end{acknowledgments}

\section{Appendix A: Couplings in the weak coupling regime ($\vert JS \vert \ll t$)}
In this appendix we calculate the magnetic exchange between localized spins in the weak coupling regime ($\vert JS \vert \ll t$). As explained in the main text, we only need to focus on the FB-FB contribution ($(0,0)$). 

The FB eigenvalues and eigenstates for both spin sector are, 
$E^{\sigma}_{0}= \pm \frac{JS}{2}=\pm \Delta$ where $+$ (resp. $-$) is for $\sigma=\uparrow$ (resp. $\sigma=\downarrow$) and
$\langle \Psi^{\sigma}_{0} \vert= (0,\frac{f_{y}}{s},-\frac{f_{x}}{s})$, where $f_{\lambda}=-2t\cos(k_{\lambda}a/2)$, with $\lambda=x,y$ and $s(\textbf{k})=\sqrt{f^{2}_x(\textbf{k}) + f^{2}_y(\textbf{k})}$.
This leads to,
\begin{eqnarray}
J^{XY}_{0,0}(\textbf{R})=-\frac{\vert JS \vert}{2} \Big|\frac{1}{N}\sum_{\textbf{k}}\dfrac{f_{XY}(\textbf{k})}{s^{2}(\textbf{k})} e^{i{\textbf{k}.\textbf{R}}} \Big|^{2},
\label{J00-appA}
\end{eqnarray}
where $f_{XY}(\textbf{k})=f^{2}_y(\textbf{k})$ for $(X,Y)=(B,B)$, $f^{2}_x(\textbf{k})$ for $(X,Y)=(C,C)$ and $f_x(\textbf{k}).f_y(\textbf{k})$ for $(X,Y)=(B,C)$. This shows that the FB-FB contribution is always ferromagnetic and it varies linearly with $\vert JS \vert$. After linearising $f_{XY}(\textbf{k})$ and $s(\textbf{k})$ in the vicinity of M($\pi$,$\pi$), Eq.(\ref{J00-appA}) becomes,
\begin{eqnarray}
\frac{1}{N}\sum_{\textbf{k}}\dfrac{f_{XY}(\textbf{k})}{s^{2}(\textbf{k})} e^{i{\textbf{k}.\textbf{R}}}=\frac{e^{i{\bf{\pi}.\textbf{R}}}}{N} \sum_{\textbf{q}} g_{XY}(\theta_{\textbf{q}})  e^{i{\textbf{q}.\textbf{R}}},
\label{J00b-appA}
\end{eqnarray}
where $g_{XY}(\theta_{\textbf{q}})=\sin^{2}(\theta_{\textbf{q}})$, $\sin(\theta_{\textbf{q}}).\cos(\theta_{\textbf{q}})$ and
$\cos^{2}(\theta_{\textbf{q}})$ for respectively $(X,Y)=(B,B)$, $(B,C)$ and $(C,C)$, with $\tan(\theta_{\textbf{q}})=\frac{q_y}{q_x}$.

First, we consider the case $(X,Y)=(B,B)$. After replacing $\frac{1}{N}\sum_{\textbf{q}} (..)$ by $\frac{a^{2}}{4\pi^{2}}\int_{0}^{q_c}q dq  \int_{-\pi}^{\pi} d\theta_{\textbf{q}} (..)$ and introducing the Bessel functions of first kind $J_{n}(u)=
\frac{i^{n}}{2\pi}\int_{-\pi}^{\pi} e^{iu\cos(\alpha)} \cos(n\alpha)d\alpha$ ($n=0,1,2,..$), one can write,
\begin{eqnarray}
\frac{1}{N}\sum_{\textbf{k}}\dfrac{f_{XY}(\textbf{k})}{s^{2}(\textbf{k})} e^{i{\textbf{k}.\textbf{R}}}= \nonumber \\
\frac{e^{i{\bf{\pi}.\textbf{R}}}}{2\pi R^{2}} \left[\cos^{2}(\theta) I_{1} + \cos(2\theta) I_2 \right],
\label{J00c-appA}
\end{eqnarray}
where $\theta$ is the angle between $\textbf{R}$ and the $x-$axis, and $I_1= \int_{0}^{\infty} J_{0}(u)udu$ and $I_2= \int_{0}^{\infty} J^{''}_{0}(u)udu$. The second derivative of $J_{0}(u)$ obey the equation  $J^{''}_{0}(u)=-J_{0}(u) + \frac{J_{1}(u)}{u}$. It can be shown that $I_1=0$ and $I_2=1$ \cite{gb-couplings,book}. We finally get, 
\begin{eqnarray}
J^{BB}_{0,0}(\textbf{R})=-\vert JS \vert \dfrac{a^{4}}{8\pi^{2}R^{4}} \cos^{2}(2\theta).
\end{eqnarray}
This result clearly explains the $1/R^{4}$ decay of the $(B,B)$ couplings found in the numerical calculations. In addition, for a fixed distance, $J^{BB}_{0,0}$, is maximum along the axes $x$ and $y$, and vanishes for $\theta=\frac{\pi}{4}$.
By symmetry arguments, one can immediately get $J^{CC}_{0,0}(\textbf{R})=J^{BB}_{0,0}(\textbf{R})$. After repeating the same kind of calculation for (B,C) pairs, one finds,
\begin{eqnarray}
J^{BC}_{0,0}(\textbf{R})=-\vert JS \vert \dfrac{a^{4}}{8\pi^{2}R^{4}} \sin^{2}(2\theta).
\end{eqnarray}

\section{Appendix B: Couplings in the strong coupling regime ($\vert JS \vert \gg t$)}

In this appendix, we calculate the magnetic couplings between localized spins in the strong coupling regime, e.g. $\vert JS \vert \gg t$. As shown in the manuscript, in this case, the couplings coincide with the $(-,+)$ contribution. Let us now derive the analytical expression of $J^{XY}_{-,+}(\textbf{R})$. 

For the spin $\uparrow$-sector, the two dispersive bands are $E^{\uparrow}_{\pm}=-g_{\mp}$ and for $\downarrow$ sector, they are $E^{\downarrow}_{\pm}=g_{\pm}$ where $g_{\pm}=\frac{1}{2} \left[-\Delta \pm \sqrt{\Delta^{2}+4s^{2}} \right]$. 
We recall that $\Delta=\frac{JS}{2}$ and $s(\textbf{k})=\sqrt{f^{2}_x(\textbf{k}) + f^{2}_y(\textbf{k})}$ (from Appendix A).
The corresponding eigenvectors are $\langle \Psi^{\uparrow}_{\pm} \vert= \frac{1}{D_{\pm}} (g_{\pm},f_{x},f_{y})$ and $\langle \Psi^{\downarrow}_{\pm} \vert= \frac{1}{D_{\mp}}
(-g_{\mp},f_{x},f_{y})$, where $D_{\pm}=\sqrt{s^{2}+ g_{\pm}^{2}}$.
From the definition of the magnetic couplings given in the main text, we get,
\begin{eqnarray}
J^{XY}_{-,+}(\textbf{R})=-\frac{(JS)^{2}}{2N^{2}}\sum_{\textbf{k},\textbf{k'}}\dfrac{h_{XY}(\textbf{k},\textbf{k'})e^{i{(\textbf{k}-\textbf{k'}).\textbf{R}}} }{(g_{+}(\textbf{k})+g_{+}(\textbf{k'})).D^{2}_{-}(\textbf{k}).D^{2}_{-}(\textbf{k'})},  \nonumber \\
\label{J+--appB}
\end{eqnarray}
where $h_{XY}(\textbf{k},\textbf{k'})=f^{2}_x(\textbf{k}).f^{2}_x(\textbf{k'})$ for $(X,Y)=(B,B)$, $h_{XY}(\textbf{k},\textbf{k'})=f^{2}_y(\textbf{k}).f^{2}_y(\textbf{k'})$ for $(X,Y)=(C,C)$ and $h_{XY}(\textbf{k},\textbf{k'})=f_x(\textbf{k}).f_y(\textbf{k}).f_x(\textbf{k'}).f_y(\textbf{k'})$ for $(X,Y)=(B,C)$.

\begin{figure}[t]\centerline
{\includegraphics[width=1\columnwidth,angle=0]{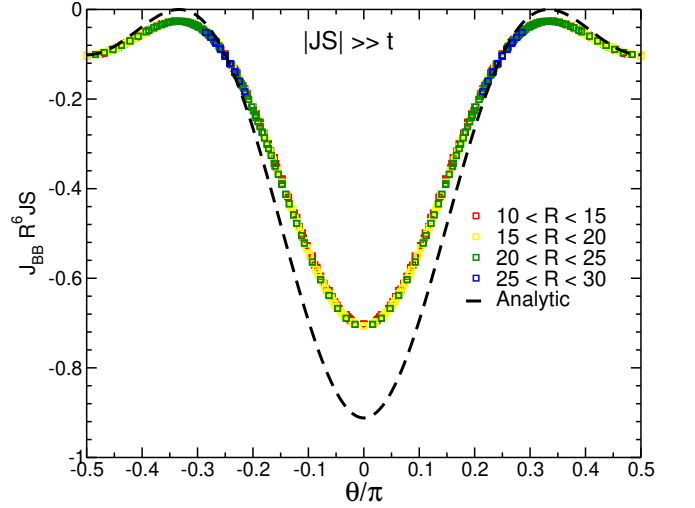}}
\vspace{-0.3cm}
\caption{(Color online) 
$\vert JS\vert.R^{6}J^{BB}$ as a function of the angle $\theta$ (angle between $\textbf{R}$ and the $x-$axis), in the limit of large JS. The symbols are the numerical calculations obtained for different distances $R$. The dashed black line is the analytical result.
} 
\label{fig1supp}
\end{figure} 

In the strong coupling regime, $\vert JS \vert \gg t$, Eq.(\ref{J+--appB}) reduces to,
\begin{eqnarray}
J^{XY}_{-,+}(\textbf{R})=-\frac{4t^{2}}{\vert JS \vert} 
\frac{1}{N^{2}}\sum_{\textbf{k},\textbf{k'}}\dfrac{h_{XY}(\textbf{k},\textbf{k'})}{s^{2}(\textbf{k})+s^{2}(\textbf{k'})} e^{i{(\textbf{k}-\textbf{k'}).\textbf{R}}}.\nonumber \\ 
\label{J+-c-appB}
\end{eqnarray}
We first consider the case $(X,Y)=(B,B)$. The linearisation in the vicinity of the M point of the Brillouin zone, leads to,
\begin{eqnarray}
J^{BB}_{-,+}(\textbf{R})= -\frac{4t^{2}}{\vert JS \vert} \frac{1}{N^{2}}\sum_{\textbf{q},\textbf{q'}} \cos^{2}(\theta_{\textbf{q}})\cos^{2}(\theta_{\textbf{q'}}) \dfrac{q^{2}q'^{2}}{q^{2}+q'^{2}}e^{i(\textbf{q}-\textbf{q'}).\textbf{R}}.
\nonumber \\
\label{J+-c-appB}
\end{eqnarray}
This equation can be rewritten, 
\begin{eqnarray}
J^{BB}_{-,+}(\textbf{R})= -\frac{a^{6} t^{2}}{\pi^{2}\vert JS \vert} \int_{0}^{q_c} \int_{0}^{q_c} \dfrac{q^{3}q'^{3}dqdq'}{q^{2}+q'^{2}} F(qR,\theta) F(q'R,\theta),\nonumber \\
\end{eqnarray}
where, $F(u,\theta)=\frac{1}{2\pi} \int_{-\pi}^{\pi} d\alpha \cos^{2}(\alpha) e^{iu\cos(\alpha-\theta)}$.
One can show that $F(u,\theta)=\cos^{2}(\theta)J_{0}(u)-\cos(2\theta) \frac{J_{1}(u)}{u}$. The Bessel functions of first kind have been defined in Appendix A. Thus, the expression of the (B,B) couplings becomes,
\begin{eqnarray}
J^{BB}_{-,+}(\textbf{R})=-\frac{a^{6}t^{2}}{\pi^{2}\vert JS \vert R^{6}}\int_{0}^{\infty} \int_{0}^{\infty} dudv\frac{u^{3}v^{3}}{u^{2}+v^{2}} \nonumber \\
\times \left[A_{\theta}J_{0}(u)- B_{\theta}\frac{J_{1}(u)}{u} \right].\left[ A_{\theta}J_{0}(v)- B_{\theta}\frac{J_{1}(v)}{v}
 \right],
 \label{jjbb-apB}
\end{eqnarray}
where we have introduced, $A_{\theta}=\cos^{2}(\theta)$ and $B_{\theta}=\cos(2\theta)$.
\begin{figure}[t]\centerline
{\includegraphics[width=1\columnwidth,angle=0]{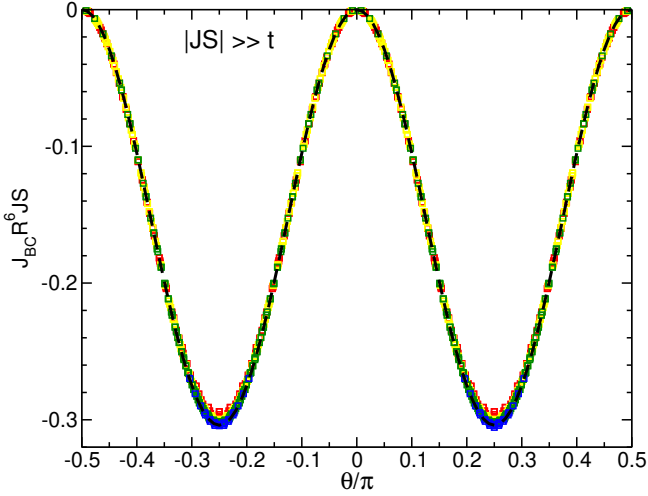}}
\vspace{-0.3cm}
\caption{(Color online) 
$\vert JS\vert  R^{6}J^{BC}$ as a function of the angle $\theta$ (angle between $\textbf{R}$ and the $x-$axis), in the limit of large $\vert JS \vert$. The symbols are the numerical calculations obtained for different distances $R$, which are the same as those of Fig.\ref{fig1supp}. The dashed line corresponds to the analytical result.
} 
\label{fig2supp}
\end{figure} 
Let us define the following set of integrals, $F^{nl}_{\mu\nu}=\int_{0}^{\infty} \int_{0}^{\infty} dudv \dfrac{u^{n}v^{l}}{u^{2}+v^{2}}J_{\mu}(u)J_{\nu}(v)$. Thus, we can rewrite Eq.(\ref{jjbb-apB}),
\begin{eqnarray}
J^{BB}_{-,+}(\textbf{R})=-\frac{a^{6}t^{2}}{\pi^{2}\vert JS \vert R^{6}}\left[A^{2}_{\theta} F_{00}^{33} + B^{2}_{\theta}F_{11}^{22} - 2A_{\theta}B_{\theta}F_{01}^{32}\right]. \nonumber \\
\label{coupbb-+}
\end{eqnarray}
The calculation of $F^{nl}_{\mu\nu}$ is detailed in the Appendix D where it has be found that
$F_{00}^{33}=4$, $F_{11}^{22}=1$ and $F_{01}^{32}=-2$. After, inserting these values in Eq.(\ref{coupbb-+}), we finally end up with,
\begin{eqnarray}
J^{BB}_{-,+}(\textbf{R})=-\frac{1}{\pi^{2}} (4\cos^{2}(\theta)-1)^{2} \frac{a^{6}}{R^{6}}\frac{t^{2}}{\vert JS \vert}.
\label{jbb-jsinfty}
\end{eqnarray}
For (B,C) pairs, one can repeat this procedure to obtain,
\begin{eqnarray}
J^{BC}_{-,+}(\textbf{R})=-\frac{3}{\pi^{2}} \sin^{2}(2\theta) \frac{a^{6}}{R^{6}} \frac{t^{2}}{\vert JS \vert}
\label{jbc-jsinfty}
\end{eqnarray}
Notice as well that $J^{CC}_{-,+}(\textbf{R})$ is straightforwardly obtained by replacing in Eq.(\ref{jbb-jsinfty}), $\theta \rightarrow \frac{\pi}{2} + \theta $.

In Fig.\ref{fig1supp} and Fig.\ref{fig2supp} are plotted for $\vert JS \vert \rightarrow \infty$, $J^{BB} \vert JS \vert R^{6}$ and $J^{BC}\vert JS \vert R^{6}$ as a function of the angle $\theta$.
First, in the case of the $(B,C)$ couplings, we observe an excellent agreement between the numerical calculations and the analytical result of Eq.(\ref{jbc-jsinfty}) for any value of $\theta$.
On the other hand, for $(B,B)$ couplings, the quality of agreement depends on $\theta$. For $\theta=\pi/2$ and $\pi/4$, the agreement is excellent, the difference is less than $1\%$. On the other hand, for $\theta=0$, the analytical result is about 28$\%$ larger than the numerical value.

\section{Appendix C: Cross-over temperature in the RPA approach}

The purpose of this appendix is to provide the analytical expression of the cross-over temperature in our two dimensional systems. For that purpose, a good tool is the Random Phase Approximation (RPA). It has been shown, in three dimensional systems, that even in disordered systems, RPA provides accurate and reliable values of the Curie temperature compared to those obtained from Monte Carlo simulation \cite{LRPA1,LRPA2}.
The Heisenberg Hamiltonian reads,
\begin{eqnarray}
H^{H}=\frac{1}{2}\sum_{i\ne j} \mathcal{J}_{ij} {\bf S}_{i}\cdot {\bf S}_{j}.
\end{eqnarray}
To start, we define the retarded Green's functions, $G^{S}_{ij,XY}(\omega) = \int^{+\infty}_{-\infty} G^{S}_{ij,XY}(t)e^{i\omega t}dt \equiv \ll S_{X,i}^{+},S_{Y,j}^{-} \gg$ where $G^{S}_{ij,XY}(t) = -i\theta(t) \bigl \langle \left[ S_{X,i}^{+},S_{Y,j}^{-}\right] \bigr\rangle$ and, $\bigl \langle ... \bigr\rangle$ denotes the thermal average.
$X$ and $Y$ are B or C. Then, we apply the equation of motion method which leads to,
\begin{eqnarray}
\omega G^{S}_{ij,BB} &=& 2\bigl \langle S_{B,i}^{z}  \bigr\rangle \delta_{ij} + \ll \left[S_{B,i}^{+},H^{H}\right],S_{B,j}^{-} \gg. \nonumber \\
\omega G^{S}_{ij,CB} &=&  \ll  \left[S_{C,i}^{+},H^{H}\right],S_{B,j}^{-} \gg.
\label{Spin-gf}
\end{eqnarray}
The terms of the form $\ll ...\gg$ which contain the commutator of spin operators with $H^{H}$ are higher order Green's functions. The RPA consists in decoupling these terms in order to get a closed form of our set of equations \cite{tyablicov}. The commutator is given by, $\left[S_{B,i}^{+},H^{H}\right]= -\sum_{Y,l} \mathcal{J}^{BY}_{il} (S_{B,i}^{+} S_{Y,l}^{z} - S_{B,i}^{z} S_{Y,l}^{+})$. The decoupling procedure is as follows,
\begin{eqnarray}
\ll \left[S_{B,i}^{+},H^{H}\right],S_{B,j}^{-} \gg &\rightarrow &  \left[-\sum_{Y,l} \mathcal{J}^{BY}_{il} \bigl \langle S^{z}\bigr\rangle \right]  G^{S}_{ij,BB} + \nonumber \\
\bigl \langle S^{z} \bigr\rangle  \sum_{Y,l} \mathcal{J}^{BY}_{il} G^{S}_{lj,YB},
\end{eqnarray}
where, for symmetry reasons, we have used the fact that $ \bigl \langle S_{B,l}^{z}\bigr\rangle = \bigl \langle S_{C,l}^{z}\bigr\rangle = \bigl\langle S^{z} \bigr\rangle$.

After Fourier transformation of Eq.(\ref{Spin-gf}), we obtain,
\begin{eqnarray}
\left[ \omega\mathbb{I} - \bigl\langle S^{z} \bigr\rangle
\begin{pmatrix}
f_{BB}^{\textbf{q}} & f_{BC}^{\textbf{q}} \\
f_{CB}^{\textbf{q}} & f_{CC}^{\textbf{q}}
\end{pmatrix}
\right]  
\begin{bmatrix}
           G_{BB}^{S}(\textbf{q},\omega)\\
           G_{CB}^{S}(\textbf{q},\omega)\\
\end{bmatrix} =
\begin{bmatrix}
           2\bigl\langle S^{z}\bigr\rangle \\
           0  \\
\end{bmatrix}, \nonumber \\
\label{matrix}
\end{eqnarray}
where, $f^{\textbf{q}}_{XX}= -h^{z}_{X}  + \bar{\mathcal{J}}^{XX}(\textbf{q}) $ with $X=B,C$.
In addition, $\bar{\mathcal{J}}^{XY}(\textbf{q})=\sum_{\textbf{R}} e^{i\textbf{q}.\textbf{R}} \mathcal{J}^{XY}(\textbf{R})$ and
$h^{z}_{X}= \sum_{Y} \bar{\mathcal{J}}^{XY}(\textbf{q}=\textbf{0})$. Finally, the off-diagonal term of the $2\times2$ matrix, $f^{\textbf{q}}_{BC}= f^{\textbf{q}}_{CB}=\bar{\mathcal{J}}^{BC}(\textbf{q})$.

The two magnons modes are straightforwardly obtained, $\omega_{\textbf{q}}^{\pm}=\bigl\langle S^{z} \bigr\rangle E_{\textbf{q}}^{\pm}$, where,
\begin{eqnarray}
E_{\textbf{q}}^{\pm}=\frac{1}{2}\left[f^{\textbf{q}}_{BB} + f^{\textbf{q}}_{CC} \pm \sqrt{(f^{\textbf{q}}_{BB}-f^{\textbf{q}}_{CC})^{2}+4(f^{\textbf{q}}_{BC}})^{2} \right]. \nonumber \\
\end{eqnarray}
The '-' sign corresponds to the acoustic mode and '+' to the optical one. It can be easily checked that in the long wave length limit ($\textbf{q} \rightarrow \textbf{0} $), $\omega_{\textbf{q}}^{-}= Dq^{2}$, where D is the spin stiffness. After inverting Eq.(\ref{matrix}), we find,
\begin{eqnarray}
G_{BB}^{S}(\textbf{q},\omega)= 2\bigl\langle S^{z} \bigr\rangle \left[\dfrac{A^{+}_{\textbf{q}}}{\omega-\omega_{\textbf{q}}^{+}}
+ \dfrac{A^{-}_{\textbf{q}}}{\omega-\omega_{\textbf{q}}^{-}}
\right],
\end{eqnarray}
where, $A^{-}_{\textbf{q}}=\dfrac{E^{+}-f_{CC}}{E^{+}-E^{-}}$ and $A^{+}_{\textbf{q}}=1-A^{-}_{\textbf{q}}$.

The magnon density of states on B sites is given by,
\begin{eqnarray}
\rho_{B}(\omega)&=&-\frac{1}{2\pi \bigl\langle S^{z} \bigr\rangle} \left[ \frac{1}{N}\sum_{\textbf{q}}G_{BB}^{S}(\textbf{q},\omega)\right]  \nonumber\\
&=&\frac{1}{N}\sum_{\textbf{q},\lambda=\pm} A^{\lambda}_{\textbf{q}}  \delta(\omega-\omega_{\textbf{q}}^{\lambda}).
\end{eqnarray}

This provides the local effective magnon occupation number at finite temperature,
\begin{eqnarray}
\Phi_{B} = \frac{1}{N}\sum_{\textbf{q}\neq 0,\lambda=\pm} \dfrac{A^{\lambda}_{\textbf{q}}}{e^{\beta\omega_{\textbf{q}}^{\lambda}}-1}. 
\end{eqnarray}
The cross-over temperature $T_{C}^{\star}$, is obtained by using the fact that when $T \rightarrow T^{\star}_C$, then $\bigl\langle S^{z} \bigr\rangle \rightarrow 0$ and $\Phi_{B} \rightarrow \infty$. From the previous equation it implies,
\begin{eqnarray}
\Phi_{B} = \frac{1}{\beta^{\star}_C}\left[ 
\frac{1}{N}\sum_{\textbf{q}\neq 0,\lambda=\pm} \dfrac{A^{\lambda}_{\textbf{q}}}{E_{\textbf{q}}^{\lambda}} \right] \frac{1}{\bigl\langle S^{z} \bigr\rangle}.
\label{eq-phi}
\end{eqnarray}

Notice as well, that at finite temperature, the local magnetization obey, for any quantum spin S, the following equation \cite{callen},
\begin{eqnarray}
\bigl\langle S^{z} \bigr\rangle=\dfrac{(S-\Phi_B)(1+\Phi_B)^{2S+1}+(S+1+\Phi_B)\Phi^{2S+1}_{B}}{(1+\Phi_B)^{2S+1}-\Phi^{2S+1}_{B}}. \nonumber \\
\label{callen}
\end{eqnarray}
Thus, in the limit where $T\rightarrow T^{\star}_C$, Eq.(\ref{callen}) becomes,
\begin{eqnarray}
\bigl\langle S^{z} \bigr\rangle=\frac{1}{3}S^{2}(1+\frac{1}{S})\frac{1}{\phi_B}.
\label{eq-phi-callen}
\end{eqnarray}
After combining Eq.(\ref{eq-phi}) and Eq.(\ref{eq-phi-callen}), we end up with the analytical expression of
$T^{\star}_C$ which reads,
\begin{eqnarray}
k_{B}T^{\star}_{C}= \frac{1}{3}(1+\frac{1}{S})\left[ 
\frac{1}{N}\sum_{\textbf{q}\neq 0,\lambda=\pm} \dfrac{A^{\lambda}_{\textbf{q}}}{S^{2}E_{\textbf{q}}^{\lambda}} \right]^{-1}.
\label{eq-Tc-bis}
\end{eqnarray}
This equation can be rewritten $k_{B}T^{\star}_{C}=(1+\frac{1}{S})k_{B}T^{\star,cl}_{C}$, where $T^{\star,cl}_{C}$ is the cross-over temperature in the case of classical spins ($S \rightarrow \infty$ with $JS$ finite).

\section{Appendix D: calculation of the integrals $F^{nl}_{\mu\nu}$.}

The purpose of this appendix is to calculate the set of integrals $F^{nl}_{\mu\nu}=\int_{0}^{\infty} \int_{0}^{\infty} dudv \dfrac{u^{n}v^{l}}{u^{2}+v^{2}}J_{\mu}(u)J_{\nu}(v)$ that are needed for the calculation of the magnetic couplings in the limit of large $\vert JS\vert$. Let us first define, for any $s \ge 0$, the following function,
\begin{eqnarray}
G^{nl}_{\mu\nu}(s)= \int_{0}^{\infty} \int_{0}^{\infty} dudv e^{-s(u^{2}+v^{2})} \dfrac{u^{n}v^{l}}{u^{2}+v^{2}}J_{\mu}(u)J_{\nu}(v). \nonumber \\
\end{eqnarray}
Then, after its derivation with respect to s, we obtain immediately,
\begin{eqnarray}
\frac{dG^{nl}_{\mu\nu}}{ds}=-T_{\mu}^{n}(s) T_{\nu}^{l}(s),
\label{gnl}
\end{eqnarray}
where we have introduced the functions $T_{\mu}^{n}(s)=\int_{0}^{\infty} x^{n}J_{\mu}(x)e^{-sx^{2}}dx$. From Ref. \cite{book} and for $\Re(\gamma) > 0$, one finds, 
\begin{eqnarray}
\int_{0}^{\infty} x^{\gamma+1}J_{\gamma}(x)e^{-sx^{2}}dx=\frac{1}{(2s)^{\gamma+1}}e^{-\frac{1}{4s}}.
\label{fonc}
\end{eqnarray}
We set $\gamma=0$ in Eq. (\ref{fonc}), and its derivation with respect to s gives,
\begin{eqnarray}
T_{0}^{3}(s)=\frac{1}{2s^{2}}(1-\frac{1}{4s})e^{-\frac{1}{4s}}.
\label{foncb}
\end{eqnarray}
In addition, if we now set $\gamma=1$ in Eq. (\ref{fonc}), then one gets,
\begin{eqnarray}
T_{1}^{2}(s)=\frac{1}{4s^{2}}e^{-\frac{1}{4s}}.
\label{foncc}
\end{eqnarray}
From Eq.(\ref{gnl}), one can write $\frac{dG^{33}_{00}}{ds}=-(T_{0}^{3}(s))^{2}$, and with the expression of $T_{0}^{3}(s)$ given in Eq.(\ref{foncb}), we obtain,
\begin{eqnarray}
G^{33}_{00}(s)= 2\int_{0}^{\frac{1}{2s}} (u^{2}-u^{3}+\frac{1}{4}u^{4})e^{-u}du,
\end{eqnarray}
where we have used the fact that $G^{33}_{00}(s=+\infty)=0$.
Finally, since $\int_{0}^{\infty} x^{n}e^{-x}dx=n!$, we immediately get,
\begin{eqnarray}
F^{33}_{00}=G^{33}_{00}(s=0)=4.
\end{eqnarray}

To calculate $F^{22}_{11}$ the procedure is similar. We now use the fact that $\frac{dG^{22}_{11}}{ds}=-(T_{1}^{2}(s))^{2}$. From Eq.(\ref{foncc}), one finds $G^{22}_{11}(s)= \frac{1}{2}\int_{0}^{\frac{1}{2s}} u^{2}e^{-u}du$. Thus,
\begin{eqnarray}
F^{22}_{11}=G^{22}_{11}(s=0)=1.
\end{eqnarray}
Similarly, we find, $G^{32}_{01}(s)= 2 \int_{0}^{\frac{1}{2s}} u^{2}(1-\frac{u}{2})e^{-u}du$, leading to,
\begin{eqnarray}
F^{32}_{01}=G^{32}_{01}(s=0)=-2.
\end{eqnarray}


\begin{thebibliography}{99}
%\bibliography{apssamp}



\bibitem{review1} D. Leykam, A. Andreanov, S. Flach, Adv Phys X,\textbf{3},1473052 (2018).
\bibitem{review2}L. Balents, C. R. Dean, D. K. Efetov and A. F. Young, Nature Physics \textbf{16}, 725(2020).


\bibitem{tang}E. Tang, J.-W. Mei, and X.-G. Wen, Phys. Rev. Lett. \textbf{106}, 236802 (2011).
\bibitem{neupert}T. Neupert, L. Santos, C. Chamon, and C. Mudry, Phys. Rev. Lett. \textbf{106}, 236804 (2011).


\bibitem{miyahara}S. Miyahara, S. Kusuta, and N. Furukawa, Physica C: Superconductivity \textbf{460,} 1145 (2007).

\bibitem{cao} Y. Cao et al., Nature \textbf{556}, 43 (2018).

\bibitem{yankowitz}M. Yankowitz et al., Science, \textbf{363}, 1059 (2019).


\bibitem{lin} Z. Lin et al., Phys. Rev. Lett.  \textbf{121}, 096401 (2018).
\bibitem{yin} J.-X. Yin, Nature Physics \textbf{15}, 443 (2019).
\bibitem{tasaki} H. Tasaki, Prog. Theor. Phys. \textbf{ 99}, 489 (1998).


\bibitem{gb-fb-paper1} G. Bouzerar and D. Mayou, Phys. Rev. Research, \textbf{2}, 033063 (2020); Phys. Rev. B, \textbf{103}, 075415 (2021).
\bibitem{mucciolo} A. Ferreira and E. R. Mucciolo, Phys. Rev. Lett. {\bf 115}, 106601 (2015). 
\bibitem{vigh} M. Vigh et al., Phys. Rev. B \textbf{88}, 161413(R) (2013).
\bibitem{qm-gb} G. Bouzerar, Phys. Rev. B \textbf{106}, 125125 (2022)



\bibitem{crgete} C. Gong, L. Li, Z. Li, et al., Nature, \textbf{546}, 265 (2017).
\bibitem{cri3} B. Huang, G. Clark, E. Navarro-Moratalla, et al., Nature \textbf{546}, 270 (2017).
\bibitem{fegete} Y. Deng, Y. Yu, Y. Song, et al., Nature, \textbf{563}, 94 (2018).
\bibitem{wang2D} Q.H. Wang,et al., ACS Nano \textbf{16}, 6960 Nano (2022).
\bibitem{yandong} M. Yandong, J. Yu, H. Thomas, 2D Mater., \textbf{4}, 025111 (2017).
\bibitem{miller}J. L. Miller, Physics Today \textbf{70}, 7,(2017).
\bibitem{gibertini} M. Gibertini, M. Koperski, A. F. Morpurgo and K. S. Novoselov 
Nature Nanotechnology \textbf{14}, 408 (2019).
\bibitem{santos} E.Santos et al., preprint, https://doi.org/10.21203/rs.3.rs-1637788/v1.







\bibitem{mermin-wagner} N.D. Mermin and H. Wagner, Phys. Rev. Lett. \textbf{17}, 1133 (1966).



\bibitem{mag2D1} H.E. Stanley, T.A. Kaplan, Phys. Rev. Lett. \textbf{17}, 913 (1966)
\bibitem{mag2D2} S.H. Shenker, and J. Tobochnik, Phys. Rev. B \textbf{22}, 4462 (1980).
\bibitem{mag2D3} H.W. J. Blote, W. Guo, H.J. Hilhorst, Phys. Rev. Lett. \textbf{88}, 047203 (2002)



\bibitem{covorc} W.Jiang, H. Huang and F. Liu Nature Comm. \textbf{10}, 2207 (2019).


\bibitem{jijcouplings} A.I. Lichtenstein, M.I. Katsnelson and V.A. Gubanov, J.
Phys.F. \textbf{14}, L125 (1984), M.I. Katsnelson and A.I. Lichtenstein, Phys. Rev. B \textbf{61}, 8906 (2000).


\bibitem{gb-couplings} G. Bouzerar, Phys. Rev. B \textbf{104}, 155151 (2021).


\bibitem{tyablicov}S.V. Tyablicov, Methods in quantum theory of magnetism
(Plenum Press, New York, 1967)


\bibitem{LRPA1} R. Bouzerar, et al., Phys. Rev. B \textbf{94}, 094437 (2016).

\bibitem{LRPA2}G. Bouzerar and O. Cépas, Phys. Rev. B \textbf{76}, 020401(R) (2007).

\bibitem{book} S. Gradshteyn and I. M. Ryzhik, Tables of Integrals, Series, and Products (Academic Press, New York, 1980),
7th edition, edited by A. Jeffrey and D. Zwillinger. 
\bibitem{callen} H.B. Callen, Phys. Rev. \textbf{130}, 890 (1963).





\end{thebibliography}
\end{document}